\documentclass[conference]{IEEEtran}
\IEEEoverridecommandlockouts
\usepackage{amsmath}
\usepackage{graphicx}
\usepackage{float}
\usepackage{amssymb}
\usepackage{algorithm}
\usepackage{algorithmic}
\usepackage{amsthm}
\usepackage{graphicx}
\usepackage{subfigure}
\usepackage{float}
\usepackage{lipsum}
\usepackage{multicol}
\usepackage[noadjust]{cite}
\def\BibTeX{{\rm B\kern-.05em{\sc i\kern-.025em b}\kern-.08em
    T\kern-.1667em\lower.7ex\hbox{E}\kern-.125emX}}

\newtheorem{theorem}{\bf Theorem}

\newtheorem{lemma}[theorem]{\bf Lemma}

\begin{document}

\title{\huge Deadline-Constrained Opportunistic Spectrum Access With Spectrum Handoff
\thanks{This work was supported in part by the National Natural Science Foundation of China under Grant 62071236 and in part by the National Science and Technology Council, Taiwan under grant MOST 110-2115-M-153-004-MY2. 
}
}
\author{\IEEEauthorblockN{Zhaolong Xue\textsuperscript{$*$}, Aoyu Gong\textsuperscript{$\flat$}, Yuan-Hsun Lo\textsuperscript{$\natural$}, Sirui Tian\textsuperscript{$*$}, and Yijin Zhang\textsuperscript{$*$}
}
    
\IEEEauthorblockA{\textsuperscript{$*$}School of Electronic and Optical Engineering, Nanjing University of Science and Technology, Nanjing 210094, China}
\IEEEauthorblockA{\textsuperscript{$\flat$}School of Computer and Communication Sciences, \'Ecole Polytechnique F\'ed\'erale de Lausanne, Lausanne 1015, Switzerland}
\IEEEauthorblockA{\textsuperscript{$\natural$}Department of Applied Mathematics, National Pingtung University, Pingtung 90003, Taiwan}

Email:
zhaolong.xue@njust.edu.cn,
aoyu.gong@epfl.ch,
yhlo@mail.nptu.edu.tw,
tiansirui@njust.edu.cn,
yijin.zhang@gmail.com
}

\maketitle

\begin{abstract}
This paper considers designing an optimal policy for deadline-constrained access in cognitive radio networks, where a secondary user needs to complete a packet transmission over the vacant spectrum within a delivery deadline.
To minimize the total access cost, it is desirable to design an optimal opportunistic access policy by utilizing channel dynamics and sensing outcomes.
We take non-negligible switching overheads, a state-dependent overtime penalty, and practical switching operations into consideration in the Markov decision process formulation of such an access problem under wide-band sensing.
Moreover, we establish the existence of monotone optimal decision rules to reduce the complexity of computing an optimal policy.
Simulation results verify our theoretical studies and the cost advantage over other policies.
\end{abstract}


\section{Introduction}
With rapid growing deadline-constrained wireless applications~\cite{Bennis2018Proc}, efficiently utilizing the scarce spectrum resources in a timely manner becomes more critical.
As a key enabler for improving the spectrum usage, cognitive radio (CR)~\cite{Mitola1999PC} was introduced to allow secondary users (SUs) to access the spectrum holes unoccupied by primary users (PUs).  
Through performing spectrum handoff~\cite{Chirstian2012CM} in CR networks (CRNs), SUs have to keep silent on unavailable channels and opportunistically switch to chosen available channels to continue their transmissions, so that the access performance of SUs can be significantly increased without harming the PUs activities. 
Obviously, waiting at an unavailable channel may lead to an overtime penalty in deadline-constrained applications~\cite{Wu2016TCOM}, while every switching to another unoccupied channel would inevitably lead to a non-negligible overhead~\cite{Demirci2020TII}.
So, it is desirable to design a policy to achieve a good tradeoff of the switching overheads and overtime penalty in spectrum handoff, by utilizing channel dynamics and sensing outcomes.

There have been many works for spectrum handoff without the deadline constraint.
Under negligible switching overheads, \cite{Zhao2007JSAC} proposed optimal and myopic policies for maximizing the average throughput built on the theory of Partially Observable Markov decision process (POMDP).
Under non-negligible switching overheads, \cite{Zhang2013JSAC} proposed optimal and myopic policies for minimizing the total service time built on the theory of Markov decision process (MDP).
Considering the throughput loss due to switching overheads, \cite{Santhoshkumar2021WCL} developed a POMDP framework to obtain an average-throughput-optimal policy with proven structural results and a near-optimal policy only based on the one-step reward.
\cite{Santhoshkumar2023NL} extended~\cite{Santhoshkumar2021WCL} to propose optimal and near-optimal policies for minimizing the average energy cost.
However, neither of~\cite{Zhao2007JSAC,Zhang2013JSAC,Santhoshkumar2021WCL,Santhoshkumar2023NL} took into account the impact of the deadline constraint on the policy design.

To deal with this impact, still under non-negligible switching overheads,~\cite{Wu2016TCOM} introduced an overtime penalty into the access cost if a data transmission has not been completed within its associated deadline, and developed an MDP framework to obtain an optimal policy with proven structural results for minimizing the total access cost.   
However,~\cite{Wu2016TCOM} ideally assumed that the idle channel set for handoff is never empty, and did not specify which target channel to handoff.
To address this impractical issue,~\cite{Ding2020TVT} developed an MDP framework with more information for channel switching, and used reinforcement learning to obtain a near-optimal access policy without requiring prior knowledge of the statistical properties of the PUs activities.
However,~\cite{Ding2020TVT} ideally assumed a constant overtime penalty which is insufficient to characterize the impact of different deadline constraint violations, and did not investigate the structure of optimal polices. 

Motivated by the aforementioned issues, this paper makes the following contributions. 
\begin{enumerate}
    \item In Section III, based on the theory of MDP, we take various practical factors (e.g. channel dynamics, switching overheads, a state-dependent overtime penalty, and practical switching operations) into consideration in the design of an optimal access policy with handoff.
    \item In Section IV, we establish the existence of monotone optimal decision rules to reduce the complexity of computing an optimal policy.
    \item In Section V, we present simulation results to verify both the structure of optimal polices and the cost advantage over other policies.
\end{enumerate}
Although the idea of using MDP in the context of spectrum handoff in CRNs is not new, our study is different because the consideration of more practical factors leads to new theoretical model properties.
The method in proving structural results is similar in some aspects to that used in~\cite{Wu2016TCOM}, but our proof is more complicated due to more complicated definitions in our MDP framework.
Due to the page limit, we have moved some of our technical proofs into our technical report~\cite{TR2023}.

\section{System Model} \label{sec2}
Consider a CRN with global synchronization, where an SU transmitter (SUTx) opportunistically sends deadline-constrained data over $M$ non-overlapping channels indexed by $\mathcal{M} \triangleq \{1,2,\cdots,M\}$. 
All the $M$ channels admit the same time-slotted structure with the same slot duration $\delta$ seconds.
The data size to be transmitted by the SUTx is $V$ bits and the associated delivery deadline is $D$ slots indexed by $\mathcal{D} \triangleq \{1,2,\cdots,D\}$. 
Let $v_t \in \mathcal{V} \triangleq \{0,1,\ldots,V\}$ denote the remaining data size at the beginning of slot $t$.

Denote by $o_t^m\in\mathcal{O}\triangleq\{0(busy),1(idle)\}$ the occupancy state of channel $m$ at slot $t$.
Let ${o}_t \triangleq[o_t^1,o_t^2,\cdots,o_t^M]$.
Assume that $o_{t}^m$ for each $m \in \mathcal{M}$ evolves independently of each other, and can be modeled by a two-state Markov chain.
Let 
\begin{align} \label{e1}
	\alpha^m=\begin{bmatrix}
		\alpha_{00}^m &\alpha_{01}^m \\ 
		\alpha_{10}^m &\alpha_{11}^m
	\end{bmatrix}
\end{align}
denote the occupancy state transition matrix of channel $m$.
The quality of each channel is described by a Gilbert-Elliot model, so that each channel over a slot may be in either a good or bad state.
Assume that the data rates per channel at the good and bad states are $R_G$ and $R_B$ bps, respectively.
Denote by $q_t^m\in\mathcal{Q}\triangleq\{0(bad),1(good)\}$ the quality state of channel $m$ at slot $t$.
Let ${q}_t \triangleq[q_t^1,q_t^2,\cdots,q_t^M]$.
Assume that $q_{t}^m$ for each $m \in \mathcal{M}$ evolves independently of each other, and can be modeled by a two-state Markov chain.
Let 
\begin{align} \label{e2}
	\beta^m=\begin{bmatrix}
		\beta_{00}^m &\beta_{01}^m \\ 
		\beta_{10}^m &\beta_{11}^m
	\end{bmatrix}
\end{align}
denote the quality state transition matrix of channel $m$.
Assume ${\alpha}^m$ and $\beta^m$ for each channel $m \in \mathcal{M}$ are both prior known to the SUTx through previous long-term channel measurements.


At each slot $t \in \mathcal{D}$, we consider different switch overheads under different scenarios. 
\begin{itemize}
    \item When the SUTx stays at the current channel to keep silent with the unfinished transmission, a silent cost $\mu_{silent}$ will be caused;
    \item When the SUTx stays at the current channel to keep silent with the finished transmission, no cost will be caused.
    \item when the SUTx switches to another channel to keep silent, a switching cost $\mu_{switch}$ and a silent cost $\mu_{silent}$ will be caused.
    \item when the SUTx switches to another channel to keep silent with the finished transmission, a switching cost $\mu_{switch}$ will be caused;
    \item when the SUTx transmits on the current channel, a transmission cost $\mu_{tr}$ will be caused.
    \item when the SUTx switches to another channel to continue its transmission, a switching cost $\mu_{switch}$, and a transmission cost $\mu_{tr}$ will be caused. 
    \item When the data has not been delivered before the specified deadline, an overtime penalty $\mu_{penalty}=w(v_{D+1})\ge0$ will be caused. Here $v_{D+1}$ denotes the remaining data size after the deadline expiration. We assume that $w(v)$ can be an arbitrary convex and nondecreasing function of $v$ with $w(0)=0$.
\end{itemize}
 The sensing overhead and switching delay are assumed small compared with other overheads and thus are ignored here~\cite{Wu2016TCOM}.



Then, at the beginning of each slot $t \in \mathcal{D}$, after performing wide-band sensing to obtain the knowledge of ${o}_t$, ${q}_t$, the SUTx makes an access decision with spectrum handoff, i.e., determine to either keep silent, transmit on the current channel, or transmit on another channel.
We want to seek a low-complexity optimal policy to minimize the sum of the expected total cost for this decision problem.

\section{Optimal Access Policies} \label{sec3}
In this section, we formulate the access problem specified in Section~II as a finite–horizon MDP, and obtain optimal policies by applying the backward induction algorithm~\cite{MDP:DSDP2005}.

\subsection{MDP Formulation}
The components of our MDP formulation are described as follows.

\underline{\emph{States:}} Define the MDP state at slot $t\in \mathcal{D}\cup \{D+1\}$ as $s_t \triangleq[v_t,{o}_t,{q}_t,c_t]$, where $v_t \in \mathcal{V}$, $o_t \in \mathcal{O}^M$, $q_t \in \mathcal{Q}^M$, and $c_t \in \mathcal{M}$ representing the index of the current channel at the beginning of slot $t$.
Denote $\mathcal{S}$ by the state space.


\underline{\emph{Actions:}}
Let $b_t$ specify the transmission decision at slot $t$, where $b_t=0$ means that the SUTx determines to keep silent while $b_t=1$ means that the SUTx determines to transmit.
Let $n_t$ specify the channel switching decision at slot $t$, where $n_t=n\in \mathcal{M}$ means that the SUTx determines to switch to channel $n$.
At the beginning of each slot $t\in \mathcal{D}$, when $s_t=s$, the SUTx performs an action $a_t=(b_t,n_t)\in \mathcal{A}_s$ where $\mathcal{A}_s$ denotes the set of allowable actions in state $s$.
Obviously, we have $\mathcal{A}_s \triangleq (\{0\} \times \mathcal{M}) \cup ( \{1\} \times \mathcal{M}_s)$ when $v_t>0$, and $\mathcal{A}_s \triangleq \{0\} \times \mathcal{M}$ when $v_t=0$.
Here $\mathcal{M}_s$ denotes the index set of the idle channels at state $s$.

\underline{\emph{Policy:}}
At the beginning of each slot $t\in \mathcal{D}$, the SUTx determines $a_t$ by an access decision function: $\pi_t: \mathcal{S} \rightarrow \mathcal{A}_s$.
An access policy is defined by a sequence of access decision functions: $\boldsymbol{\pi}\triangleq(\pi_1,\pi_2,\cdots,\pi_D)$.
Let $\boldsymbol{\Pi}$ denote the set of all possible such policies.

\underline{\emph{Cost Function:}}
We define the cost at state $s=(v,o,q,c)$ with action $a=(b,n)\in \mathcal{A}_s$ at slot $t \in \mathcal{D}$ as
\begin{align} \label{e3}
     & h_t(s,a) = h_t(v,o,q,c,b,n) \notag \\
     & \quad\quad =
     \begin{cases} 
        \mu_{silent}, & \text{ if } b=0,n=c,v>0, \\
        \mu_{silent}+\mu_{switch}, & \text{ if } b=0,n\neq c,v>0, \\
        \mu_{tr}, & \text{ if }  b=1,n=c, v>0,\\
        \mu_{tr}+\mu_{switch}, & \text{ if } b=1,n \neq c,v>0,\\
        \mu_{switch}, & \text{ if } b=0,n \neq c,v=0,\\
        0, & \text{ otherwise}. 
    \end{cases}
\end{align}
The cost at slot $D+1$ is the overtime penalty $w(v_{D+1})$.

\underline{\emph{State Transition Function:}}
The state transition function $\chi_{s',s,a}$ is defined as the transition probability of moving from the state $s_t=s$ to $s_{t+1}=s'$ when the SUTx performs action $a_t=a \in \mathcal{A}_s$ at the beginning of slot $t\in \mathcal{D}$.
So, we have 
\begin{align} \label{e4}
	& \chi_{s',s,a} = \chi_{(v',o',q',c'), (v,o,q,c), (b,n)} = \text{Pr}(o_{t+1}=o'|o_t=o) \nonumber \\
  & \cdot \text{Pr}(q_{t+1}=q'|q_t=q) \cdot \text{Pr}(c_{t+1}=c'|c_t=c,n_t=n)     \nonumber   \\	
  & \cdot\text{Pr}(v_{t+1}=v' | v_t=v,o_t=o,q_t=q,b_t=b,n_t=n).
\end{align}
where $\text{Pr}(o_{t+1}=o'|o_t=o)$ can be obtained from (1), $\text{Pr}(q_{t+1}=q'|q_t=q)$ can be obtained from (2), $\text{Pr}(c_{t+1}=c'|c_t=c,n_t=n)=1$ if $c'=n$, and 
\begin{align} \label{e5}
	&\text{Pr}(v_{t+1}=v' | v_t=v,o_t=o,q_t=q,b_t=b,n_t=n) \nonumber \\
	&\quad\quad= \begin{cases} 1, & \text{ if }v'=\big[v-r(o,q,b,n)\big]^+,\\
		0, &\text{ otherwise. }
	\end{cases}
\end{align}
Here $[x]^+=\max\{0,x\}$ and $r(o,q,b,n)$ denotes the data rate function.
We have
\begin{align} \label{e6}
    r(o,q,b,n) =
    \begin{cases} 
        R_G \delta, & \text{ if } o^n=1,q^n=1,b=1, \\
		R_B \delta, & \text{ if }  o^n=1,q^n=0,b=1, \\
             0,                &   \text{ otherwise}.
    \end{cases}
\end{align}
We assume that $r(o,q,b,n)$ always takes an integral value. 

\subsection{MDP Solution}
We aim to find an optimal policy $\boldsymbol{\pi}^*$ that minimizes the expected total access cost from slot 1 to $D+1$, i.e.,
\begin{align} \label{e7}
    & {\boldsymbol{\pi}}^* \in \arg\min_{\boldsymbol{\pi}\in\boldsymbol{\Pi}} \notag \\
    & \quad \mathbb{E}^{\boldsymbol{\pi}}\left[ \sum_{t=1}^D h_t\big(s_t,\pi_t(s_t)\big) + w(v_{D+1}) | s_1=(V,o,q,c) \right].
\end{align}

Let $U_{t}^*(s)$ denote the minimum cost from slot $t$ to $D+1$ when $s_{t}=s$.
So, by \eqref{e3}--\eqref{e6}, we have the following recursive equations:
\begin{align} \label{e9}
	U^*_{D+1}(s)&= w(v_{D+1}), \quad  \forall s \in\mathcal{S},  \\ \label{e10}
	U^*_{t}(s)&=\min_{a\in \mathcal{A}_s}  U_t(s,a), \quad \forall s \in\mathcal{S}, \forall t \in\mathcal{D},
\end{align}
where 
\begin{align} \label{e8}
	& U_t(s,a) = h_t(s,a) + \sum_{s'\in\mathcal{S}} \chi_{s',s,a} \cdot U^*_{t+1}(s'). 
\end{align}
Applying the backward induction algorithm~\cite{MDP:DSDP2005} to the above can lead to ${\boldsymbol{\pi}}^*$.

Finally we discuss the monotone property of $U_t^*(s)$ that will be useful in proving monotone optimal decision rules.
\begin{lemma} \label{l1}
	$U_t^*(v,o,q,c)$ is nondecreasing in $v$ for each $t\in\mathcal{D}\cup\{D+1\}$, $o \in\mathcal{O}^M$, $q\in\mathcal{Q}^M$, and $c \in\mathcal{M}$.
\end{lemma}
The proof of Lemma~1 is provided in Appendix A.

\section{Monotone Optimal Decision Rules} \label{sec4}
To reduce the computational complexity to obtain an optimal policy, this section aims to establish the existence of monotone optimal decision rules and propose a monotone backward induction algorithm.


We begin with introducing the definitions of superadditivity and subadditivity~\cite{SAP1998}.
Let $X$ and $Y$ be partially ordered sets and $g(x,y)$ be a real-valued function on $X\times Y$.
Let $\overline{x}$, $\underline{x}$ denote the elements in $X$, respectively, while $\overline{y}$, $\underline{y}$ are the same in $Y$.
Then $g$ is said to be superadditive if
\begin{align} \label{e11}
	g(\overline{x},\overline{y}) + g(\underline{x},\underline{y}) \geq g(\overline{x},\underline{y}) + g(\underline{x},\overline{y}),
\end{align}
for $\overline{x} \geq \underline{x}$ in $X$ and $\overline{y} \geq \underline{y}$ in $Y$.
On the other hand, $g$ is said to be subadditive if the reverse inequality above holds.

\begin{lemma}[\cite{MDP:DSDP2005}, Lemma~4.7.1, Ch.~4] \label{l2}
	If $g(x,y)$ is a superadditive function on $X\times Y$ and $\max_{y\in Y} g(x,y)$ exists for each $x\in X$, $f(x)\triangleq \text{max} \{\mathop{\arg\max}_{y\in Y} g(x,y)\}$ is monotone nondecreasing in $x$.
\end{lemma}

We need the following subadditive property of $U_t(s,a)$ to prove monotone optimal decision rules.
The proof is provided in our technical report~\cite{TR2023}.


\begin{lemma} \label{newl3}
    Define a partial order in $\mathcal{A}_s$: $(\overline{b},\overline{n})\geq(\underline{b},\underline{n})$ if $r(o,q,\overline{b},\overline{n})\geq r(o,q,\underline{b},\underline{n})$ for arbitrary $(\overline{b},\overline{n}), (\underline{b},\underline{n}) \in\mathcal{A}_s$.
Then, $U_t(v,o,q,c,a)$ is subadditive on $\mathcal{V} \times \mathcal{A}_s$ for each $t\in\mathcal{D}$, $o\in\mathcal{O}^M$, $q\in\mathcal{Q}^M$ and $c\in\mathcal{M}$.
\end{lemma}


Now we provide conditions which ensure the optimality of monotone decision rules through utilizing Lemmas~\ref{l1}--\ref{newl3}. 

\begin{theorem} \label{t5}
Under an arbitrary current state $s=(v,o,q,c) \in \mathcal{S}$, there exists monotone optimal decision rules in the following cases.
    \begin{enumerate}
    \item  When the current channel is idle and has the best quality in $\mathcal{M}_s$, i.e., $o^{c}=1$ and $q^c=\mathop{\max}_{m\in\mathcal{M}_{s}}q^m$, we have
    \begin{align} \label{e12}
    	\pi_t^*(s)=\begin{cases}(0,c), & \text{ if } v < V_t^{th1}(o,q,c), \\
    		(1,c), & \text{ otherwise}, \end{cases} 
    \end{align}
     for each $t \in\mathcal{D}$ and each possible $s$ in this case.
    \item When the current channel is busy and there exists an idle channel $n^{\diamond}$ that has the best quality in $\mathcal{M}_s$, i.e., $o^{c}=0$, $\exists n^{\diamond} \in\{q^i=\mathop{\max}_{m\in\mathcal{M}_{s}}q^m | i\in\mathcal{M}_{s}\}$, we have
    \begin{align} \label{e13}
    	\pi_t^*(s)=\begin{cases} (0,c), & \text{ if } v < V_t^{th2}(o,q,c), \\
    		(1,n^{\diamond}), & \text{ otherwise}, \end{cases}
    \end{align}
    for each $t \in\mathcal{D}$ and each possible $s$ in this case.
    \item When the current channel is idle and bad but there exists an idle and good channel $n^\star$, i.e., $o^c=1$, $q^c=0$, $\exists n^\star  \in\{q^i=1| i\in\mathcal{M}_{s}\}$, we have
    \begin{align} \label{e14}
    	\pi_t^*(s)=\begin{cases} (0,c), & \text{ if } v < V_t^{th3}(o,q,c), \\
            (1,c), & \text{ if }  V_t^{th3}(o,q,c) \leq v < V_t^{th4}(o,q,c), \\
    		(1,n^\star), & \text{ otherwise}, \end{cases}
    \end{align}
    for each $t \in\mathcal{D}$ and each possible $s$ in this case.
    \item When all the channels are busy, i.e., $\mathcal{M}_s=\emptyset$, we have
    \begin{equation} \label{e15}
        \pi_t^*(s) = (0,c), 
    \end{equation}
     for each $t \in\mathcal{D}$ and each possible $s$ in this case.
    \end{enumerate}
The threshold $V_t^{th1}(o,q,c)$, $V_t^{th2}(o,q,c)$, $V_t^{th3}(o,q,c)$, and $V_t^{th4}(o,q,c)$ can be obtained by Algorithm~\ref{algo1}.
\end{theorem}

The proof of Theorem~\ref{t5} is given in Appendix B.


\begin{algorithm}[ht] 
    \caption{A monotone backward induction algorithm for finding ${\boldsymbol{\pi}}^*$} \label{algo1}
    \renewcommand{\algorithmicrequire}{\textbf{Input:}}
    \renewcommand{\algorithmicensure}{\textbf{Output:}}
    \begin{algorithmic}[1]
        \REQUIRE The sampling interval $\zeta=1$ in $\mathcal{V}$; The size of data to be transmitted $V$; The constrained deadline $D$; The overtime penalty $U_{D+1}^*(s)$ for $\forall s \in\mathcal{S}$;
        \ENSURE An optimal policy ${\boldsymbol{\pi}}^*$.
        \STATE $t=D$
        \WHILE{$t>0$}
        \FOR{$o\in\mathcal{O}^M$ and $q\in\mathcal{Q}^M$}
        \FOR{$c=1$ to $M$}
        \IF{$\mathcal{M}_s=\emptyset$}
        \STATE $\pi_t^*(s)=(0,c)$; continue;
        \ENDIF
        \STATE $v=0$
        \WHILE{$v\leq V$}
        \STATE Calculate $U_t(s,a)$ by \eqref{e8}
        \STATE $\pi_t^*(s) \in \arg\min_{a\in\mathcal{A}_s} U_t(s,a)$
        \STATE $U_t^*(s)=U_t(s,\pi_t^*(s))$
        \IF{$\pi_t^*(s)=(1,c)$}
        \IF{$s$ satisfies case 1 in Theorem~\ref{t5}}
        \STATE $V_t^{th1}(o,q,c)=v$
        \FOR{$v=V_t^{th1}(o,q,c)+1$ to $V$}
        \STATE $\pi_t^*(s)=(1,c)$
        \ENDFOR
        \ELSE
        \STATE $V_t^{th3}(o,q,c)=v$; continue;
        \ENDIF
        \ELSIF{$\pi_t^*(s)\neq (0,c)$ or $(1,c)$}
        \IF{$s$ satisfies case 2 in Theorem~\ref{t5}}
        \STATE $V_t^{th2}(o,q,c)=v$
        \FOR{$v=V_t^{th2}(o,q,c)+1$ to $V$}
        \STATE $\pi_t^*(s)=(1,n^\diamond)$
        \ENDFOR
        \ELSE
        \STATE $V_t^{th4}(o,q,c)=v$
        \FOR{$v=V_t^{th4}(o,q,c)+1$ to $V$}
        \STATE $\pi_t^*(s)=(1,n^\star)$
        \ENDFOR
        \ENDIF
        \ENDIF
        \STATE $v=v+\zeta$
        \ENDWHILE
        \ENDFOR
        \ENDFOR
            \STATE $t=t-1$
        \ENDWHILE
        \RETURN An optimal policy ${\boldsymbol{\pi}}^*$.
    \end{algorithmic}
\end{algorithm}



Different from the backward induction algorithm~\cite{MDP:DSDP2005}, Algorithm~1 simplifies finding an optimal policy into determining the threshold for some cases.
Note that increasing the sampling interval $\zeta$ in $\mathcal{V}$ can further reduce the computational complexity of Algorithm~1 but sacrifices the optimality.

\section{Results} \label{sec5}
This section illustrates the structural results of the proposed optimal policy as proved in Theorem 5, and compares the total access costs of the proposed optimal policy, the always-staying policy (i.e., the SUTx always stays on its current channel), and the quality-based-switching policy (i.e., the SUTx always selects the nearest idle channel with the best quality).

The scenarios considered in the numerical experiments are in accordance with the descriptions in
Section~\ref{sec2}.
For the channel related parameters, we set $M=3$, $\alpha^m=[0.2,0.8;0.8,0.2]$ for each channel $m$, $\beta^m=[0.5,0.5;0.5,0.5]$ for each channel $m$, $\delta=1$ second, $R_G=2$M bps, and $R_B=1$M bps.
Further, for costs under different scenarios, we set $\mu_{silent}=0.01$, $\mu_{tr}=40$, $\mu_{switch}=5$, and $\mu_{penalty}=Lv^2$ with $L=5 \times 10^{-12}$.
We shall vary other network configurations over a wide range to investigate the impact of policy design on the total access cost.

\subsection{Optimality of Monotone Policies}
\begin{figure*}[ht!]
\centering
\subfigure[] {
    \includegraphics[width=0.23\textwidth]{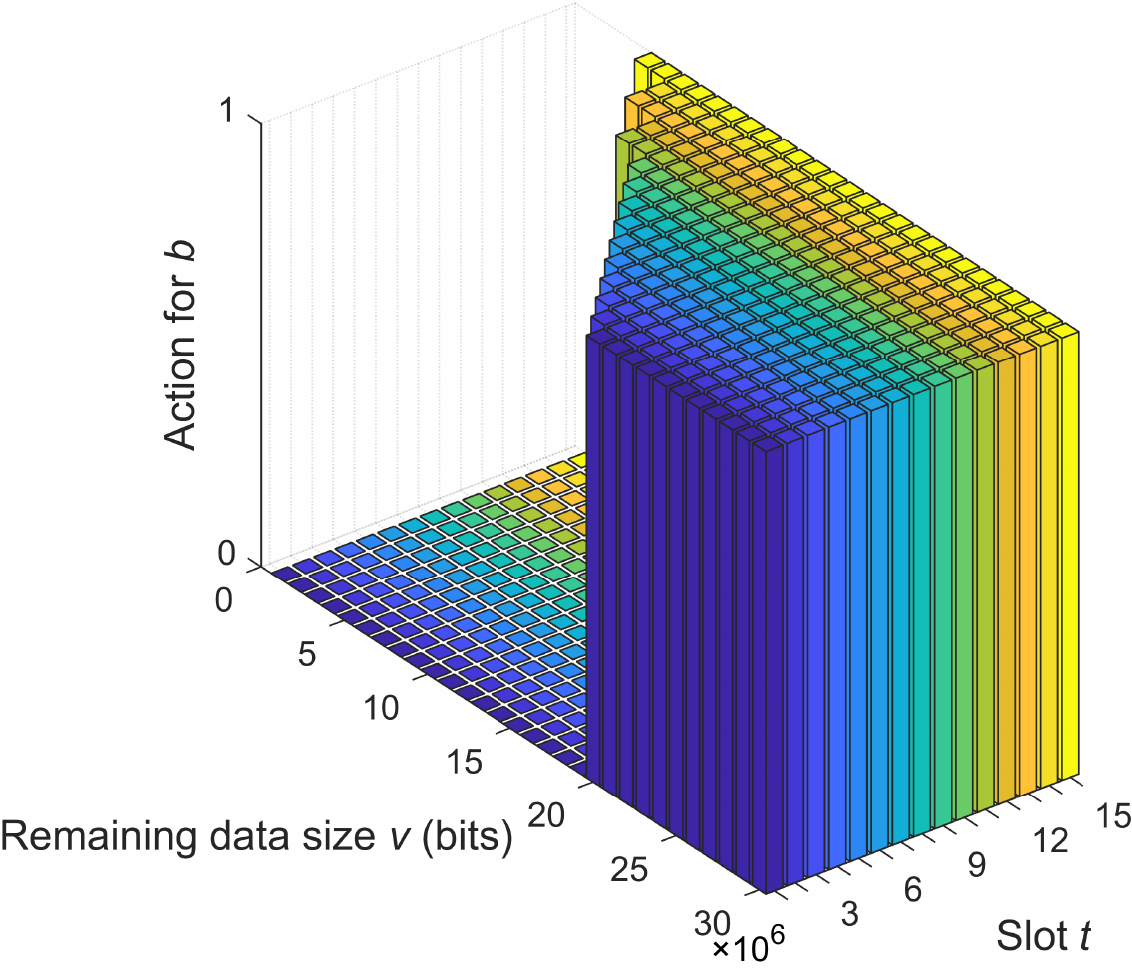}
}
\subfigure[] {
    \includegraphics[width=0.23\textwidth]{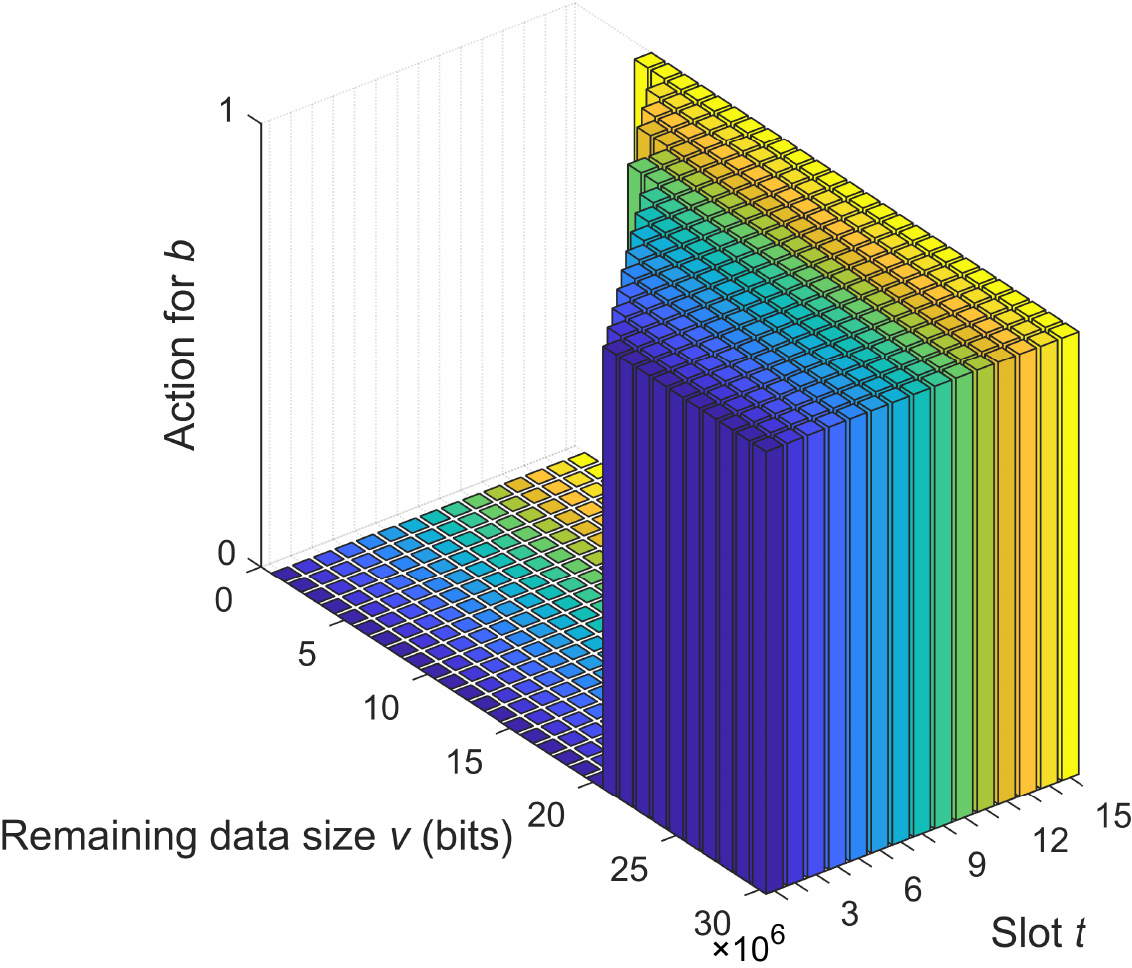}
}
\subfigure[] {
    \includegraphics[width=0.23\textwidth]{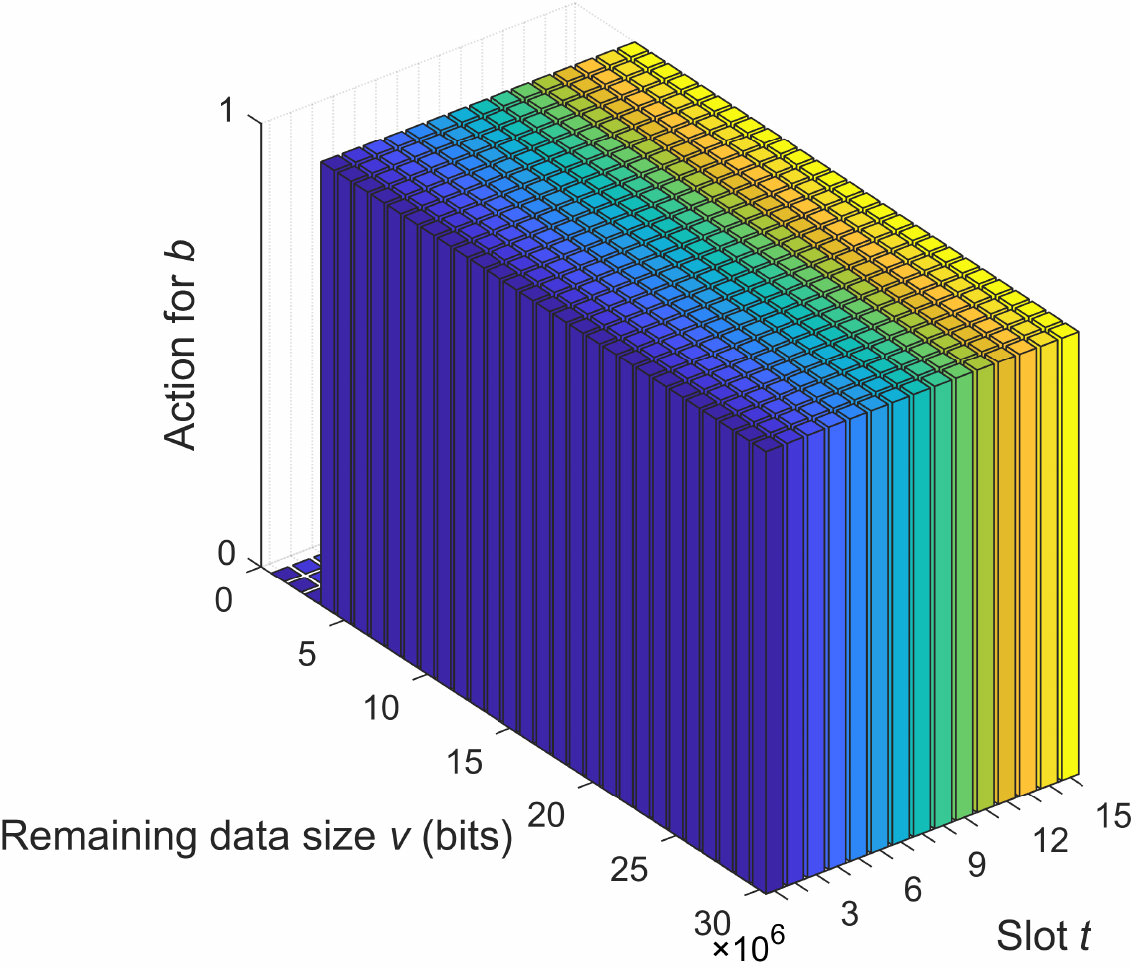}
}
\subfigure[] {
    \includegraphics[width=0.23\textwidth]{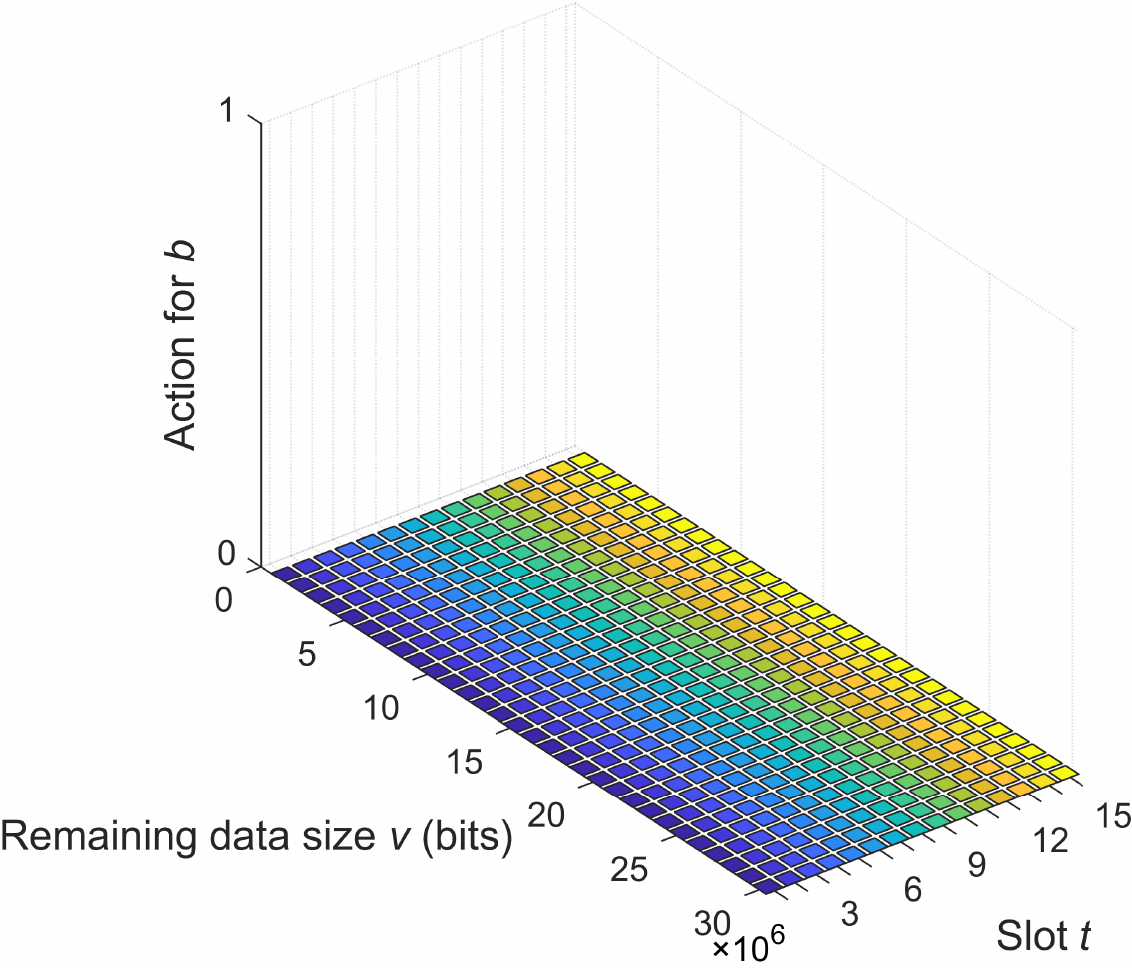}
}
\subfigure[] {
    \includegraphics[width=0.23\textwidth]{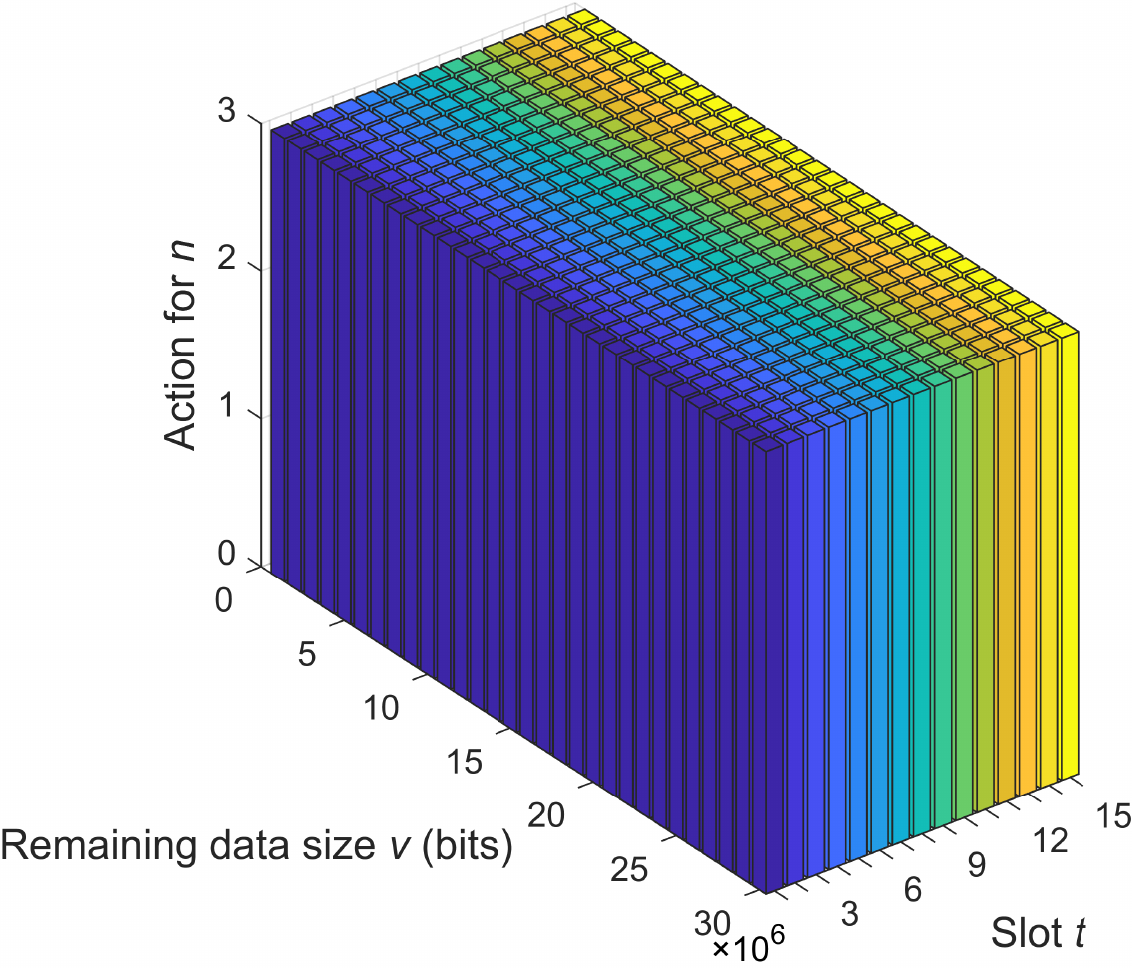}
}
\subfigure[] {
    \includegraphics[width=0.23\textwidth]{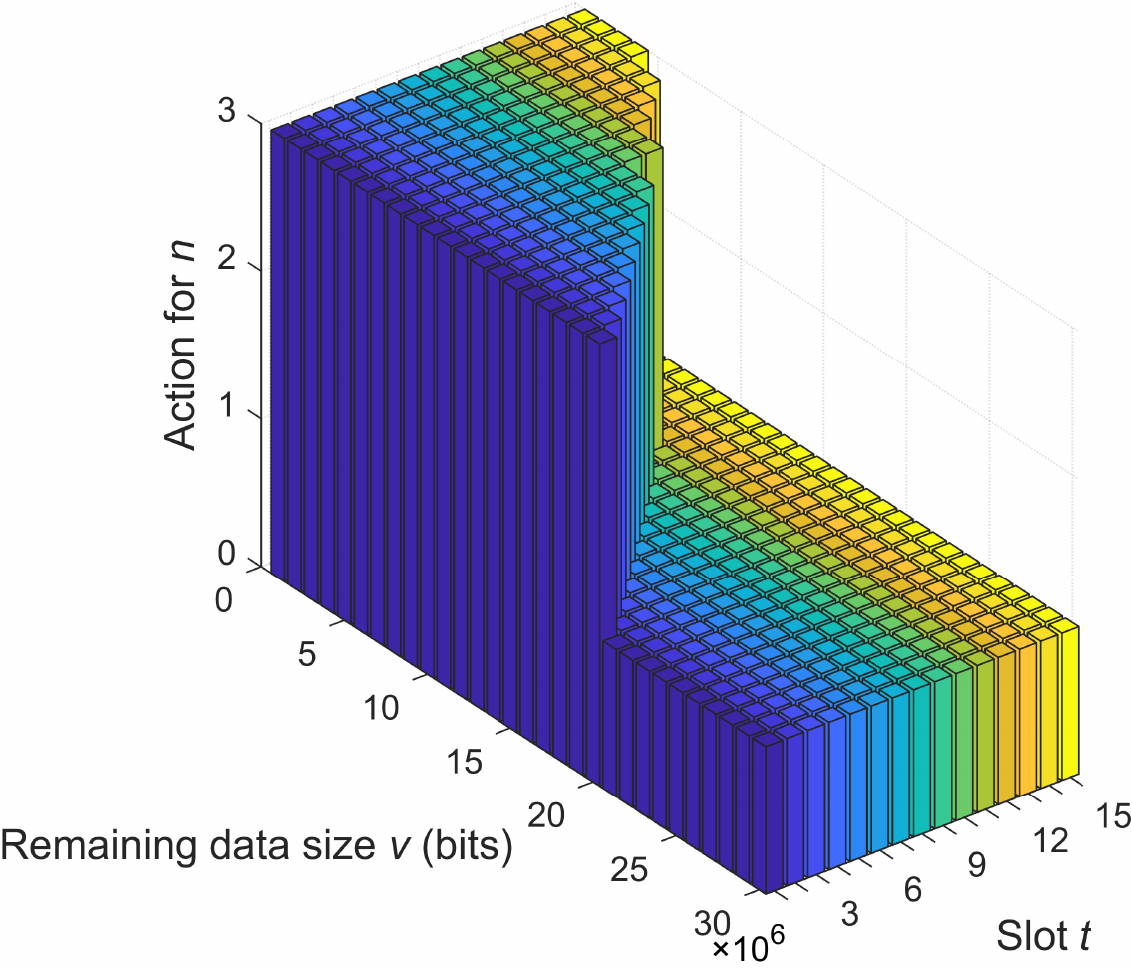}
}
\subfigure[] {
    \includegraphics[width=0.23\textwidth]{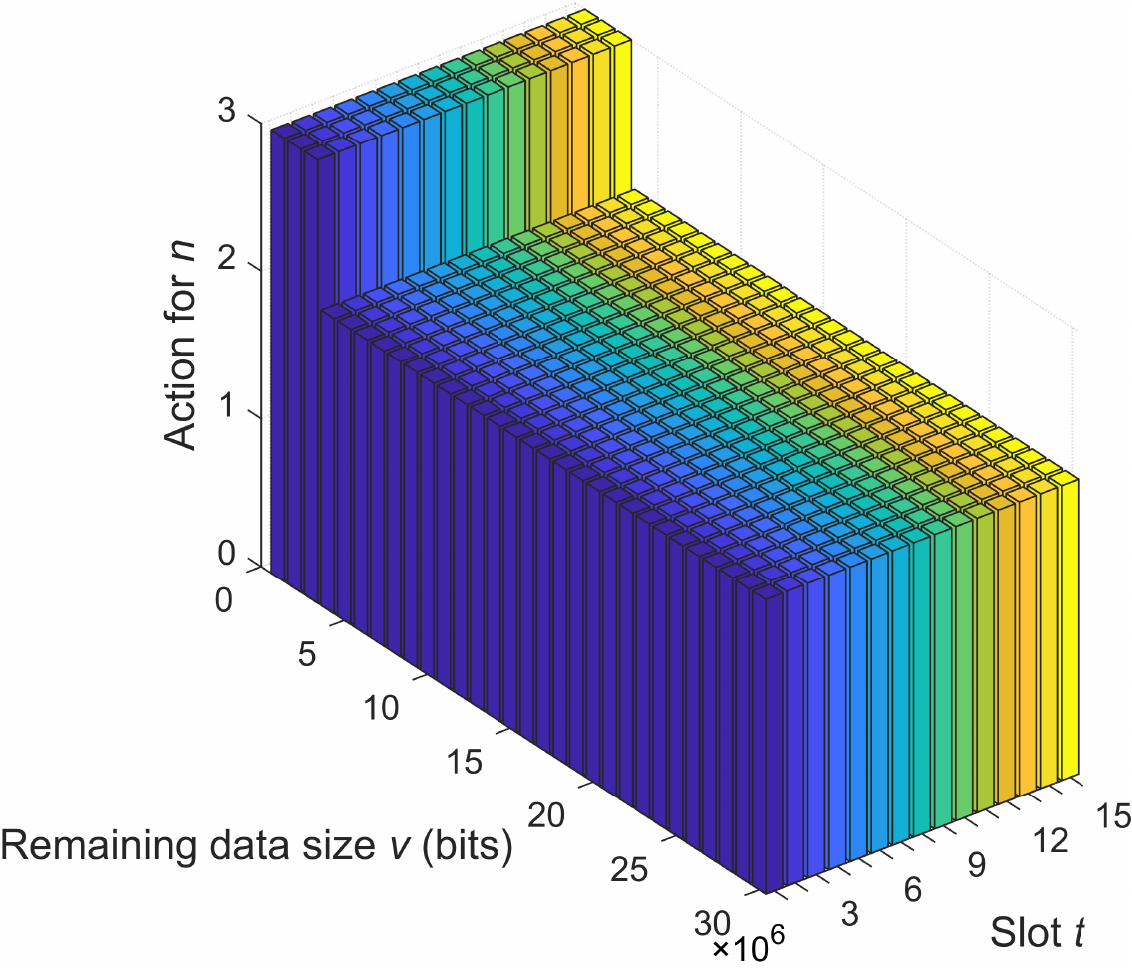}
}
\subfigure[] {
    \includegraphics[width=0.23\textwidth]{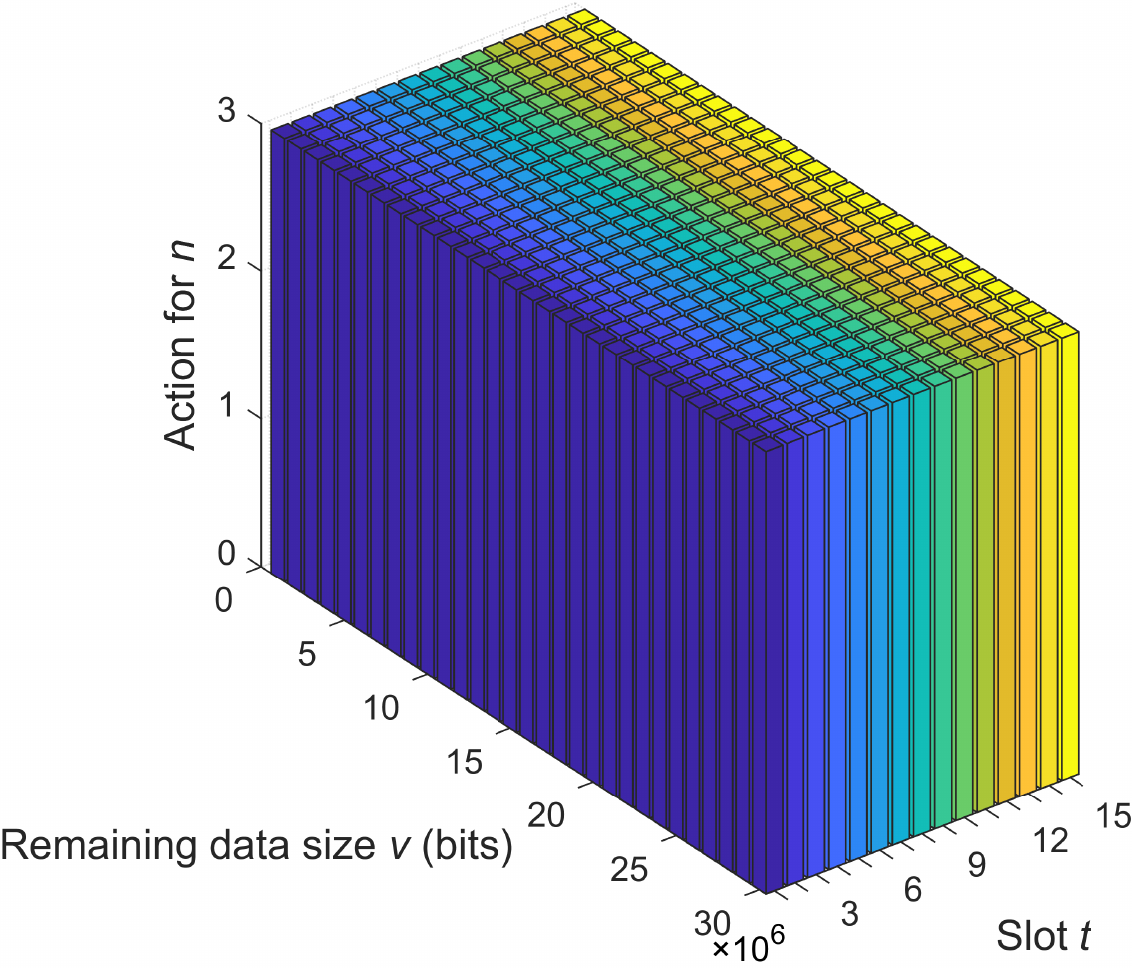}
}
\caption{Optimal actions for (a)(e)~case~1 in Theorem~\ref{t5}, (b)(f)~case~2 in Theorem~\ref{t5}, (c)(g)~case~3 in Theorem~\ref{t5}, and (d)(h)~case~4 in Theorem~\ref{t5} when $V=30$M, $D=15$, $q=[0,1,0]$, and $c=3$.} 
\label{fig:action space}
\end{figure*}

Fig.~\ref{fig:action space} shows optimal actions obtained from the backward induction algorithm corresponding to the four cases in Theorem~\ref{t5}.
The $x$-axis represents the current slot $t$, the $y$-axis represents the remaining data size $v$, and
the $z$-axis represents the optimal action $a$.
We observe that the optimal actions are indeed of the threshold structure in $v$.
For case~1 in Theorem~5, Fig.~\ref{fig:action space}(a)(e) shows that the SUTx decides to transmit in order to decrease the overtime penalty if $v \geq V_t^{th1}(o,q,c)$, but decides to keep silent otherwise.
The reason is that such a decision would lead to a smaller cost than spectrum switching.
For case~2, Fig.~\ref{fig:action space}(b)(f) shows that the SUTx decides to switch to an idle channel with the best quality to transmit when $v \geq V_t^{th2}(o,q,c)$, since the increment of total access cost caused by the overtime penalty is much greater than that caused by spectrum switching.
For case~3, Fig.~\ref{fig:action space}(c)(g) shows that the SUTx decides to switch to an idle and good channel to continue its transmission rather than transmit on the current channel if $v \geq V_t^{th4}(o,q,c)$, but decides to keep silent otherwise.
Here $V_t^{th3}(o,q,c)=V_t^{th4}(o,q,c)$. 
This is because the reduction of $v$ caused by a higher channel rate would lead to a lower overtime penalty, which is greater than the switching cost.
For case~4, Fig.~\ref{fig:action space}(d)(h) shows that the SUTx decides to always keep silent at the current channel, due to switching to another channel would lead to a switching cost without any benefit. 

\subsection{Comparisons With Other Policies}

\begin{figure}[ht!]
	\centering
		\includegraphics[width=3.2in]{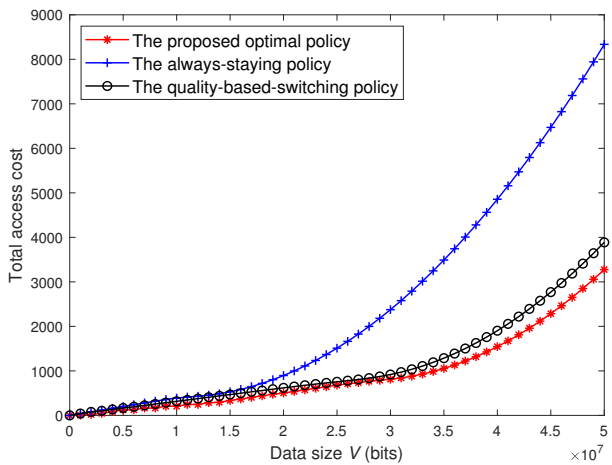}
	\caption{Total access cost versus the data size when $D=20$, $o_1=[1, 1, 1]$, and $q_1=[0, 1, 0]$.}
	\label{fig:total cost with fixed deadline}
\end{figure}

Fig.~\ref{fig:total cost with fixed deadline} shows the total access cost versus the data size $V$.
We observe that the proposed optimal policy obtains the lowest total access cost for all the cases.
This is because that the SUTx under the always-staying policy has to keep silent when the current channel is busy, resulting in a larger penalty for unfinished transmission than other polices, while the SUTx under the quality-based-switching policy makes more frequent spectrum switching resulting in more switching cost.
In addition, we observe that the cost advantage of the proposed optimal policy increases with $V$.
The reason is that the impact of the overtime penalty on the total access cost increases with $V$, then the optimal policy plays a more important role in reducing the cost.



\begin{figure}[htbp]
	\centering
		\includegraphics[width=3.2in]{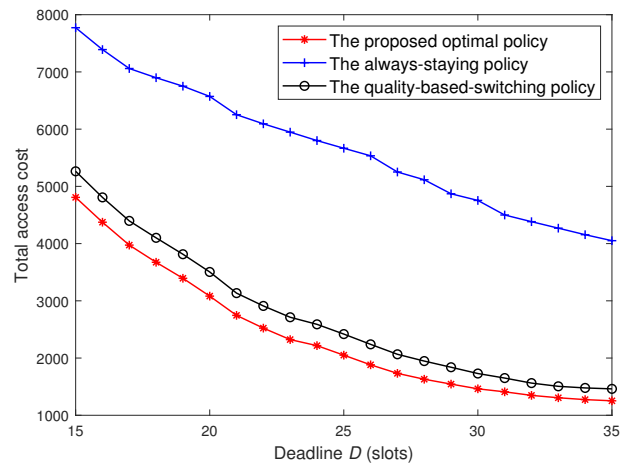}
	\caption{Total access cost versus the delivery deadline when $V=50$M, $o_1=[1, 1, 1]$, and $q_1=[0, 1, 0]$.}
	\label{fig:deadline cost}
\end{figure}

Fig.~\ref{fig:deadline cost} shows the total access cost versus the delivery deadline $D$.
We observe that the proposed optimal policy still obtains the lowest total access cost for all the cases.
In addition, we observe that the cost advantage of the proposed optimal policy decreases with $D$.
The reason is that the SUTx has more chances to complete the data transmission, which weakens the effect of policy design.


\section{Conclusion} \label{sec6}
In this paper, we have investigated the deadline-constrained spectrum access problem in CRNs by taking into account various practical factors in spectrum handoff.
We formulated such an access problem under wide-band sensing as a finite-horizon MDP and proposed an optimal policy that achieves an optimal tradeoff of switching overheads and overtime penalty, thereby minimizing the total access cost.
Further, we proved the existence of monotone optimal decision rules to reduce the complexity of computing an optimal policy.
Simulation results verify our theoretical studies and show that the proposed optimal policy outperforms other policies.
Our future work is to use the theory of POMDP to formulate this access problem under narrow-band sensing and investigate the threshold-structure optimal policies. 

\appendices
\section{Proof of Lemma~\ref{l1}}
	We shall prove $U_t^*(v,o,q,c)$ is nondecreasing in $v$ by induction on $t$ from $t=D+1$ down to $1$.
	As the penalty function $w(v)$ is nondecreasing in $v$, we know $U_{D+1}^*(s)=w(v_{D+1})$ is nondecreasing in $v$.
	Next, when $t\in \mathcal{D}$, we assume $U_{t+1}^*(v,o,q,c)$ is nondecreasing in $v$.
	Since $h_t(s,a)$ is independent of $v$ by~\eqref{e3} and $\chi_{s',s,a} \geq 0$ by~\eqref{e4}, we obtain $U_t^*(v,o,q,c)$ is nondecreasing in $v$. 
	Since both the base case and the inductive step have been proved as true, we have completed the proof.

\section{Proof of Theorem~\ref{t5}}
By Lemmas~\ref{l2} and~\ref{newl3}, we know that $\pi_t^*(v,o,q,c)$ is nondecreasing in $v$, 
We shall apply this property to prove the monotone decision rules in the following four cases.

\emph{Case 1:} Consider that the current channel is idle and has the best quality in $\mathcal{M}_s$, i.e., $o^{c}=1$ and $q^c=\mathop{\max}_{n\in\mathcal{M}_{s}}q^n$.
Obviously, we have $r(o,q,b,c) \geq r(o,q,b,n)$ for arbitrary $n\in\mathcal{M}_s$, $b\in\{0,1\}$.
By this property together with \eqref{e3}, \eqref{e6}, and Lemma~\ref{l1}, we further have
\begin{align} \label{t5_opt}
    & U_t(v,o,q,c,b,c) \notag \\
    & = h_t(v,o,q,c,b,c) + \sum_{o'\in\mathcal{O}^M} \sum_{q'\in\mathcal{Q}^M} \text{Pr}(o_{t+1}=o'|o_t=o) \nonumber \\
    & \quad \cdot \text{Pr}(q_{t+1}=q'|q_t=q) U_{t+1}^* \big( [v-r(o,q,b,c)]^+,o',q',c \big) \nonumber \\
    & \leq h_t(v,o,q,c,b,n) + \sum_{o'\in\mathcal{O}^M} \sum_{q'\in\mathcal{Q}^M} \text{Pr}(o_{t+1}=o'|o_t=o) \nonumber \\
    & \quad \cdot \text{Pr}(q_{t+1}=q'|q_t=q) U_{t+1}^* \big( [v-r(o,q,b,n)]^+,o',q',n \big) \nonumber \\
    & = U_t(v,o,q,c,b,n),
\end{align}
for arbitrary $n\in\mathcal{M}_s$, $b\in\{0,1\}$.
So, by~\eqref{t5_opt}, it is optimal for the SUTx to choose $(b,c)$ with an optimal value of $b$. 
On the other hand, since $\pi_t^*(v,o,q,c)$ is nondecreasing in $v$, when to choose $(1,c)$ or $(0,c)$ can be determined by the threshold $V_t^{th1}(o,q,c)$ as shown in Algorithm 1.



\emph{Case 2:} Consider that the current channel is busy and there exists an idle channel $n$ that has the best quality in $\mathcal{M}_s$, i.e., $o^{c}=0$, $\exists n^\diamond \in\{q^i=\mathop{\max}_{m\in\mathcal{M}_{s}}q^m | i\in\mathcal{M}_{s}\}$. 
By \eqref{e3}, \eqref{e6}, and Lemma~\ref{l1}, we have
\begin{align} \label{t5_opt2}
     U_t(v,o,q,c,0,c) < U_t(v,o,q,c,0,n),
\end{align}
for arbitrary $n\in\mathcal{M}$, and 
\begin{align} \label{t5_opt3}
     U_t(v,o,q,c,1,n^\diamond) \leq U_t(v,o,q,c,1,n),
\end{align}
for arbitrary $n\in\mathcal{M}_s$.
So, by~\eqref{t5_opt2} and~\eqref{t5_opt3}, it is optimal for the SUTx to choose an optimal action from $(0,c)$ and $(1,n^\diamond)$.    
Further, since $\pi_t^*(v,o,q,c)$ is nondecreasing in $v$, when to choose $(0,c)$ or $(1,n^\diamond)$ can be determined by the threshold $V_t^{th2}(o,q,c)$ as shown in Algorithm 1.

\emph{Case 3:} Consider that the current channel is idle and bad but there exists an idle and good channel $n^\star$, i.e., $o^c=1$, $q^c=0$, $\exists n^\star\in\{q^i=1| i\in\mathcal{M}_{s}\}$.
By~\eqref{e3} and~\eqref{e6}, we have
\begin{equation}
    h_t(v,o,q,c,1,c) < h_t(v,o,q,c,1,n^\star),
\end{equation}
\begin{equation}
    r(o,q,1,c) < r(o,q,1,n^\star).
\end{equation}
The SUTx can transmit more data when choosing the action $(1,n^\star)$ at the cost of a switching cost compared to choosing $(1,c)$.
So, by~\eqref{t5_opt2}, it is optimal for the SUTx to choose an optimal action from $(0,c)$, $(1,c)$ and $(1,n^\star)$.
Further, since $\pi_t^*(v,o,q,c)$ is nondecreasing in $v$, when to choose $(0,c)$, $(1,c)$ or $(1,n^\star)$ can be determined by the thresholds $V_t^{th3}(o,q,c)$ and $V_t^{th4}(o,q,c)$ as shown in Algorithm 1.

\emph{Case 4:} Consider that all the channels are busy, i.e., $\mathcal{M}_s=\emptyset$.
So, we have $\mathcal{A}_s\triangleq\{0\}\times \{\mathcal{M}\}$.
By~\eqref{t5_opt2}, we know that the SUTx would always prefer to choose the action $(0,c)$ rather than $(0,n)$ for arbitrary $n\in \mathcal{M}$.

\onecolumn
{
\centering
{\LARGE Supplemental Material for the paper ``Deadline-Constrained} \par
}
\vspace{0.25cm}

{
\centering
{\LARGE Opportunistic Spectrum Access With Spectrum Handoff''} \par
}
\vspace{0.5cm}

{
\centering
{\large Zhaolong Xue, Aoyu Gong, Yuan-Hsun Lo, Sirui Tian, and Yijin Zhang} \par
}
\vspace{0.5cm}

\section*{Proof of Lemma~\ref{newl3}}
Before proving the subadditivity of $U_t(v,o,q,c,a)$, we should prove the inequality 
\begin{align} \label{cond_lemma3}
    & U_t^*\big([v-r(o,q,\underline{b},\underline{n})]^+,o,q,\underline{n}\big) - U_t^*\big([v-r(o,q,\overline{b},\overline{n})]^+,o,q,\overline{n}\big) \nonumber  \\
    & \quad \geq U_t^*\big([v-r(o,q,\underline{b},\underline{n})-\varepsilon]^+,o,q,\underline{n}\big) - U_t^*\big([v-r(o,q,\overline{b},\overline{n})-\varepsilon]^+,o,q,\overline{n}\big),
\end{align}
holds for each $t\in\mathcal{D}\cup\{D+1\}$, $v\in\mathcal{V}$, $o\in\mathcal{O}^M$, $q\in\mathcal{Q}^M$, and an arbitrary non-negative integer $\varepsilon$.
We prove it by mathematical induction.
First, since the penalty function $w(v)$ is convex and nondecreasing in $v$, by the property of convex function, we have
\begin{align}
    & w\big([v-r(o,q,\underline{b},\underline{n})]^+\big) - w\big([v-r(o,q,\overline{b},\overline{n})]^+\big) \geq w\big([v-r(o,q,\underline{b},\underline{n})-\varepsilon]^+\big) - w\big([v-r(o,q,\overline{b},\overline{n})-\varepsilon]^+\big).
\end{align}
Then we obtain
\begin{align}
    & U_{D+1}^*\big([v-r(o,q,\underline{b},\underline{n})]^+,o,q,\underline{n}\big) - U_{D+1}^*\big([v-r(o,q,\overline{b},\overline{n})]^+,o,q,\overline{n}\big) \nonumber  \\
    & \quad = w\big([v-r(o,q,\underline{b},\underline{n})]^+\big) - w\big([v-r(o,q,\overline{b},\overline{n})]^+\big) \nonumber \\
    & \quad \geq w\big([v-r(o,q,\underline{b},\underline{n})-\varepsilon]^+\big) - w\big([v-r(o,q,\overline{b},\overline{n})-\varepsilon]^+\big) \nonumber \\
    & \quad = U_{D+1}^*\big([v-r(o,q,\underline{b},\underline{n})-\varepsilon]^+,o,q,\underline{n}\big) - U_{D+1}^*\big([v-r(o,q,\overline{b},\overline{n})-\varepsilon]^+,o,q,\overline{n}\big).
\end{align}
Next, when $t\in\mathcal{D}$, we assume that
\begin{align}
    & U_{t+1}^*\big([v-r(o,q,\underline{b},\underline{n})]^+,o,q,\underline{n}\big) - U_{t+1}^*\big([v-r(o,q,\overline{b},\overline{n})]^+,o,q,\overline{n}\big) \nonumber  \\
    & \quad \geq U_{t+1}^*\big([v-r(o,q,\underline{b},\underline{n})-\varepsilon]^+,o,q,\underline{n}\big) - U_{t+1}^*\big([v-r(o,q,\overline{b},\overline{n})-\varepsilon]^+,o,q,\overline{n}\big).
\end{align}
From \eqref{e10}, let
\begin{align} \label{cond1}
    U_t^*\big([v-r(o,q,\underline{b},\underline{n})]^+,o,q,\underline{n}\big) & = U_t\big([v-r(o,q,\underline{b},\underline{n})]^+,o,q,\underline{n},d_1\big),  \\ 
    U_t^*\big([v-r(o,q,\overline{b},\overline{n})]^+,o,q,\overline{n}\big) & = U_t\big([v-r(o,q,\overline{b},\overline{n})]^+,o,q,\overline{n},d_2\big),  \\ 
    U_t^*\big([v-r(o,q,\underline{b},\underline{n})-\varepsilon]^+,o,q,\underline{n}\big) & = U_t\big([v-r(o,q,\underline{b},\underline{n})-\varepsilon]^+,o,q,\underline{n},d_3\big),  \\ 
    U_t^*\big([v-r(o,q,\overline{b},\overline{n})-\varepsilon]^+,o,q,\overline{n}\big) & = U_t\big([v-r(o,q,\overline{b},\overline{n})-\varepsilon]^+,o,q,\overline{n},d_4\big), \label{cond4}
\end{align}
where $d_1,d_2,d_3,d_4\in\mathcal{A}_s$. So we have 
\begin{align} 
    & U_t^*\big([v-r(o,q,\underline{b},\underline{n})]^+,o,q,\underline{n}\big) - U_t^*\big([v-r(o,q,\overline{b},\overline{n})]^+,o,q,\overline{n}\big) \nonumber  \\
    &  \quad\quad - U_t^*\big([v-r(o,q,\underline{b},\underline{n})-\varepsilon]^+,o,q,\underline{n}\big) + U_t^*\big([v-r(o,q,\overline{b},\overline{n})-\varepsilon]^+,o,q,\overline{n}\big) \nonumber \\
    &  = U_t\big([v-r(o,q,\underline{b},\underline{n})]^+,o,q,\underline{n},d_1\big) - U_t\big([v-r(o,q,\overline{b},\overline{n})]^+,o,q,\overline{n},d_2\big) \nonumber \\
    &  \quad\quad - U_t\big([v-r(o,q,\underline{b},\underline{n})-\varepsilon]^+,o,q,\underline{n},d_3\big) + U_t\big([v-r(o,q,\overline{b},\overline{n})-\varepsilon]^+,o,q,\overline{n},d_4\big) \nonumber \\
    &  = \Big(\underbrace{U_t\big([v-r(o,q,\underline{b},\underline{n})]^+,o,q,\underline{n},d_1\big) - 
        U_t\big([v-r(o,q,\underline{b},\underline{n})-\varepsilon]^+,o,q,\underline{n},d_1\big)}_{X_1}\Big) \nonumber \\
    &  \quad\quad + \Big(\underbrace{U_t\big([v-r(o,q,\underline{b},\underline{n})-\varepsilon]^+,o,q,\underline{n},d_1\big) - U_t\big([v-r(o,q,\underline{b},\underline{n})-\varepsilon]^+,o,q,\underline{n},d_3\big)}_{X_2}\Big) \nonumber \\
    &  \quad\quad + \Big(\underbrace{-U_t\big([v - r(o,q,\overline{b},\overline{n})]^+,o,q,\overline{n},d_2\big) + U_t\big([v -     r(o,q,\overline{b},\overline{n})]^+,o,q,\overline{n},d_4\big)}_{X_3}\Big) \nonumber \\
    &  \quad\quad - \Big(\underbrace{U_t\big([v -     r(o,q,\overline{b},\overline{n})]^+,o,q,\overline{n},d_4\big) - U_t\big([v-r(o,q,\overline{b},\overline{n})-\varepsilon]^+,o,q,\overline{n},d_4\big)}_{X_4}\Big). \nonumber
\end{align}
Here, we get $X_2, X_3\geq 0$ by~\eqref{cond1}--\eqref{cond4}.
Then, by \eqref{e8} and induction hypothesis, for $d_4=(b',n')$, we have
\begin{align}
    X_4 & = \sum_{o'\in\mathcal{O}^M} \sum_{q'\in\mathcal{Q}^M} \text{Pr}(o_{t+1}=o'|o_t=o) \cdot \text{Pr}(q_{t+1}=q'|q_t=q) \notag \\
    & \quad \cdot \Big[ U_{t+1}^*\big([v-r(o,q,\overline{b},\overline{n}) - r(o,q,b',n')]^+,o',q',n'\big)  - U_{t+1}^*\big([v-r(o,q,\overline{b},\overline{n}) - r(o,q,b',n') - \varepsilon]^+,o',q',n'\big) \Big] \nonumber \\
    & \leq \sum_{o'\in\mathcal{O}^M} \sum_{q'\in\mathcal{Q}^M} \text{Pr}(o_{t+1}=o'|o_t=o) \cdot \text{Pr}(q_{t+1}=q'|q_t=q) \notag \\
    & \quad \cdot \Big[ U_{t+1}^*\big([v - r(o,q,\underline{b},\underline{n}) - r(o,q,\overline{b},\overline{n})]^+,o',q',\overline{n}\big) - U_{t+1}^*\big([v - r(o,q,\underline{b},\underline{n}) - r(o,q,\overline{b},\overline{n}) -\varepsilon]^+,o',q',\overline{n}\big) \Big]. \nonumber
\end{align}
Similarly, we can obtain
\begin{align}
    X_1 & \geq \sum_{o'\in\mathcal{O}^M} \sum_{q'\in\mathcal{Q}^M} \text{Pr}(o_{t+1}=o'|o_t=o) \cdot \text{Pr}(q_{t+1}=q'|q_t=q) \notag \\
    & \quad \cdot \Big[ U_{t+1}^*\big([v - r(o,q,\underline{b},\underline{n}) - r(o,q,\overline{b},\overline{n})]^+,o',q',\overline{n}\big) - U_{t+1}^*\big([v - r(o,q,\underline{b},\underline{n}) - r(o,q,\overline{b},\overline{n}) -\varepsilon]^+,o',q',\overline{n}\big) \Big]. \nonumber
\end{align}
Hence, $X_1\geq X_4$.
Thus, the inequality \eqref{cond_lemma3} holds.

For $\forall \overline{v},\underline{v} \in \mathcal{V}$, $t\in\mathcal{D}$, let $\underline{v}=[\overline{v}-\gamma\zeta]^+$, where $\gamma$ is a positive integer and $\zeta=1$ is the sampling interval in $\mathcal{V}$, from \eqref{cond_lemma3}, we have
\begin{align} \label{cond_subadd}
    & U_{t+1}^*([\overline{v}-r(o,q,\underline{b},\underline{n})]^+,o,q,\underline{n}) - U_t^*([\overline{v} - r(o,q,\overline{b},\overline{n})]^+,o,q,\overline{n}) \nonumber \\
    & \quad \geq U_{t+1}^*([\overline{v} - r(o,q,\underline{b},\underline{n}) - \zeta]^+,o,q,\underline{n}) - U_t^*([\overline{v}-r(o,q,\overline{b},\overline{n}) - \zeta]^+,o,q,\overline{n}) \nonumber \\
    & \quad \geq U_{t+1}^*([\overline{v} - r(o,q,\underline{b},\underline{n}) - \gamma\zeta]^+,o,q,\underline{n}) - U_t^*([\overline{v} -r(o,q,\overline{b},\overline{n}) - \gamma\zeta]^+,o,q,\overline{n}) \nonumber \\
    & \quad = U_{t+1}^*([\underline{v} - r(o,q,\underline{b},\underline{n})]^+,o,q,\underline{n}) - U_t^*([\underline{v} - r(o,q,\overline{b},\overline{n})]^+,o,q,\overline{n}).
\end{align}
Thus, for $d_1=(\overline{b},\overline{n})\geq d_2=(\underline{b},\underline{n})$, by \eqref{cond_subadd}, we have
\begin{align}
    & U_t(\overline{v},o,q,c,d_1) - U_t(\overline{v},o,q,c,d_2) - U_t(\underline{v},o,q,c,d_1) + U_t(\underline{v},o,q,c,d_2) \nonumber \\
    & = \sum_{o'\in\mathcal{O}^M} \sum_{q'\in\mathcal{Q}^M} \text{Pr}(o_{t+1}=o'|o_t=o) \cdot \text{Pr}(q_{t+1}=q'|q_t=q)  \nonumber \\
    & \quad\quad \cdot \Big[ U_{t+1}^*\big([\overline{v} - r(o,q,\overline{b},\overline{n})]^+,o',q',\overline{n}\big) - U_{t+1}^*\big([\underline{v} - r(o,q,\overline{b},\overline{n})]^+,o',q',\overline{n}\big) \Big] \nonumber \\
    & \quad\quad - \sum_{o'\in\mathcal{O}^M} \sum_{q'\in\mathcal{Q}^M} \text{Pr}(o_{t+1}=o'|o_t=o) \cdot \text{Pr}(q_{t+1}=q'|q_t=q)  \nonumber \\
    & \quad\quad\quad\quad \cdot \Big[ U_{t+1}^*\big([\overline{v} - r(o,q,\underline{b},\underline{n})]^+,o',q',\underline{n}\big) - U_{t+1}^*\big([\underline{v} - r(o,q,\underline{b},\underline{n})]^+,o',q',\underline{n}\big) \Big]  \nonumber \\
    &\leq 0.
\end{align}
From the definition of subadditivity, we conclude that $U_t(v,o,q,c,a)$ is subadditive on $\mathcal{V} \times \mathcal{A}_s$.

\end{document}